\documentclass[oribibl]{llncs}
\usepackage{makeidx,graphicx,calc,booktabs,tabularx,amsmath,layout,framed}  

\makeatletter
\newcommand\footnoteref[1]{\protected@xdef\@thefnmark{\ref{#1}}\@footnotemark}
\makeatother
\usepackage{dcolumn}
\usepackage{color}
\graphicspath{{figure/}}
\newcolumntype{R}{>{\raggedleft\arraybackslash}X}
\newcolumntype{C}{>{\centering\arraybackslash}X}
\newcolumntype{L}{>{\raggedright\arraybackslash}X}
\newcolumntype{Y}{D..{6.4}}

\begin{document}

\title{Assessing Levels of Attention\\
using Low Cost Eye Tracking
}

\author{Per B{\ae}kgaard\thanks{Acknowledgment: This work is supported in part by the Innovation Fund Denmark through the project Eye Tracking for Mobile Devices.} \and Michael Kai Petersen \and Jakob Eg Larsen}

\institute{
Cognitive Systems\\
Department of Applied Mathematics and Computer Science\\
Technical University of Denmark, Building 321\\
DK-2800 Kgs.Lyngby, Denmark\\
\email{\{pgba,mkai,jaeg\}@dtu.dk}
}

\maketitle

\begin{abstract}


The emergence of mobile eye trackers embedded in next generation smartphones or VR displays will make it possible to trace not only what objects we look at but also the level of attention in a given situation. Exploring whether we can quantify the engagement of a user interacting with a laptop, we apply mobile eye tracking in an in-depth study over 2 weeks with nearly 10.000 observations to assess pupil size changes, related to attentional aspects of alertness, orientation and conflict resolution. Visually presenting conflicting cues and targets we hypothesize that it's feasible to measure the allocated effort when responding to confusing stimuli.
Although such experiments are normally carried out in a lab, we are able to differentiate between sustained alertness and complex decision making even with low cost eye tracking ``in the wild''. From a quantified self perspective of individual behavioral adaptation, the correlations between the pupil size and the task dependent reaction time and error rates may longer term provide a foundation for modifying smartphone content and interaction to the users’ perceived level of attention.



\begin{keywords}
Eye Tracking, Attention Network
\end{keywords}
\end{abstract}

{
  \begin{center}
  \begin{framed}
  \textit{
    \small
    \noindent This is an author-generated preprint.\\
    To be published in the HCI International 2016 Conference Proceedings.\\
    The final publication will be  available at Springer via http://dx.doi.org/TODO
  }
  \end{framed}
  \end{center}
}

\section{Introduction}

Low cost eye trackers which can be embedded in next generation smartphones will enable design of cognitive interfaces that adapt to the user’s perceived level of attention. 
Even when ``in the wild", and no longer constrained to fixed lab setups, mobile eye tracking provides novel opportunities for continuous self-tracking of our ability to perform a variety of tasks across a number of different contexts.

Interacting with a smartphone screen requires attention which in turn involves different networks in the brain related to alertness, spatial orientation and conflict resolution~\cite{posner2012attentional}. 
These aspects can be separated by flanker-type of experiments with differently cued, sometimes conflicting, prompts.
Dependent on whether the task involves fixating the eyes on an unexpected part of the screen, or resolving the direction of an arrow surrounded by distracting stimuli, different parts of the attention network will be activated, in turn resulting in varying reaction times~\cite{fan2002testing}.

The dilation and constriction of the pupil is not only triggered by changes in light and fixation but reflect fluctuations in arousal networks in the brain ~\cite{joshi2016relationships}, which from a quantified self perspective may enable us to assess whether we are sufficiently concentrated when we interact with the screens of smartphones or laptops, carrying out our daily tasks. Likewise the pupil size increases when we face an unexpected uncertainty ~\cite{ang2015commentary}, physically apply force by flexing muscles, or motivationally have to decide on whether the outcome of a task justifies the required effort ~\cite{varazzani2015noradrenaline}. Thus, when we perform specific actions, the cognitive load involved can be estimated using eye tracking. The pupil dilates if the task requires a shift from a sustained tonic alertness and orientation to more complex decision making, in turn triggering a phasic component caused by the release of norepinephrine neurotransmitters in the brain~\cite{aston-jones2005integrative},~\cite{gabay2011orienting}, which may reflect both the increased energization as well as the unexpected uncertainty related to the task ~\cite{ang2015commentary}.


Whereas these results have typically been obtained under controlled lab conditions, we explore in the present study the feasibility of assessing a user’s level of attention ``in the wild" using mobile eye tracking.

\section{Method}

\subsection{Experimental Procedure}

This longitudinal study was performed repeatedly over the course of two weeks in September-October 2015. Two male right-handed subjects, A and B, (of average age 56) each performed a session very similar to the Attention Network Test (\textsc{ant})~\cite{fan2002testing} approximately twice every weekday, resulting in 16 resp. 17 complete datasets, totaling $9.504$ individual reaction time tests. The experiment ran ``in the wild" in typical office environments off a conventional MacBook Pro 13" (2013 model with Retina screen) that had an Eye Tribe Eye Tracker connected to it. The \textsc{ant} used here is implemented in PsychoPy~\cite{peirce2007psychopy} and is available on github~\cite{baekgaard2016ant}. Simultaneously, eye tracking data is recorded at 60 Hz and timestamped for synchronization through the Eye Tracker API~\cite{eyetribe2015api} via the PeyeTribe~\cite{baekgaard2015peyetribe} interface.

\begin{figure}[h!]
\def\svgwidth{\textwidth}
{\sffamily
\small
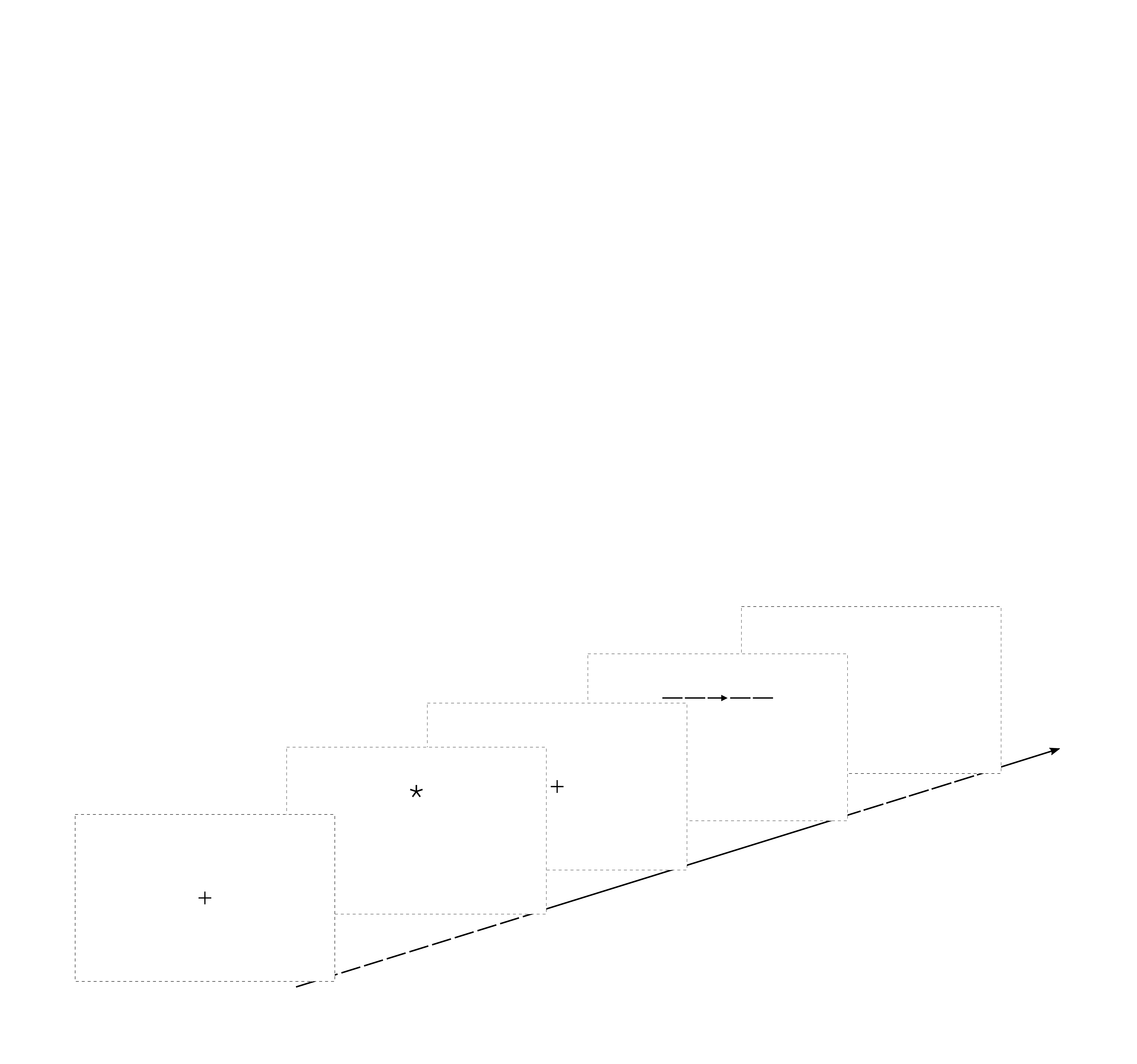
}
\caption[LoF Entry]{
This Attention Network Test procedure used here: Every 4 seconds, a cue (either of 4 conditions (\textsc{Top, Left})) precedes a target (either of 3 congruency conditions (\textsc{Top, Right})), to which the participant responds by pressing a key according to the central arrow. 
The reaction time differences between cue- and congruency conditions form the basis for calculating the latencies of the attention, orientation and conflict resolution networks.
}
\label{fig:ant-run-exp}
\end{figure}

Before the actual experimental procedure starts, a calibration of the Eye Tracker is performed. The experiment contains an initial trial run that the user may select to abort, after which 3 rounds of $2\cdot48$ conditioned reaction time tests follows (Fig. \ref{fig:ant-run-exp}); each test is conditioned on one of 3 targets: \emph{Incongruent}, \emph{Neutral} or \emph{Congruent} and on 4 cues: \emph{No Cue}, \emph{Center Cue}, \emph{Double Cue} or \emph{Spatial Cue}. At the start of each test, a fixation cross appears, and after a random delay of $0.4-1.6$s the user is presented to a cue (when present for the particular condition). $0.5$s later the target appears, either with incongruent, neutral or congruent flankers. The user is instructed to hit a button on the left or right side of the keyboard with his left or right hand depending on the direction of the central arrow of the target, which appeared above or below the initial centred fixation cross. Half the targets appear above and half below the fixation cross, and left/right pointing central arrows also appear evenly distributed. The resulting reaction time ``from target presentation to first registered keypress" is logged, together with the conditions of the individual tests, whether the user hit the correct left/right key or not, and a common timestamp. For further details on the \textsc{ant} please see~\cite{fan2002testing}.
 
Each test takes approximately $4$s to perform. With $2\cdot3$ repetitions of all combinations of conditions, left/right arrows and above/below targets, this results in $6\cdot12\cdot2\cdot2=288$ single tests. The user has the option of a short break after each 96 performed tests. A typical session with calibration, experimental procedure and short breaks lasts approximately 25-30 minutes.

\subsection{Analysis}

The reaction times for each experiment, for which the user responded correctly within $1.7$s, are grouped and averaged over each of the $3$ congruency and $4$ cue conditions, and the Attention Network Test timings can be calculated as follows:
\begin{align*}
t_{\operatorname{alertness}} &= \overline{t_{\operatorname{no~cue}}} - \overline{t_{\operatorname{double~cue}}} \\
t_{\operatorname{orientation}} &= \overline{t_{\operatorname{center~cue}}} - \overline{t_{\operatorname{spatial~cue}}} \\
t_{\operatorname{conflict~resolution}} &= \overline{t_{\operatorname{incongruent}}} - \overline{t_{\operatorname{congruent}}} \\
\end{align*}
where
$$
\overline{t_{\operatorname{cond}}} =  {1 \over N} {\sum_{i|i = \operatorname{cond}}^N{t_i}}
$$

Linear pupil size and inter-pupil distance data can be somewhat ``noisy'' when recording in office conditions. After epoch'ing to corresponding cue times for the individual tests, invalid/missing data from blink-affected periods are removed, and a Hampel~\cite{hampel1974influence} filter is therefore applied, using a centered window of $\pm83$ms (shorter than a typical blink) and a limit of $3\sigma$, to remove remaining outliers. Data is then downsampled to $100$ms resolution using a windowed averaging filter, and scaled proportionally to the value at epoch start (cue presentation), so that the resulting pupil dilations represent relative change\footnote{The data received from the eye tracker is uncalibrated and cannot easily be referenced to a metric measurement.} vs the pupil size at cue presentation. This last part was done to compensate for varying environmental luminosity changes and, to some degree, to offset any effect from immediately preceding reaction time test(s) and to compensate for accidental head position drift.

Time-locked averaging is then done by grouping data from similar conditions within each experiment, from which the group-mean relative pupil dilations can be derived.

At the same time, the inter-pupil distance is calculated, to ensure that pupil size changes would not be the accidental result of moving the head slightly during the experiment. Additionally, a ``baseline" experiment has been performed, recording eye tracking data in a condition where no action can be taken by the user and when no arrow-heads are visible on the targets but otherwise presented in similar conditions, in order to rule out that the recorded pupil dilations would be the result of (small) luminosity changes caused by the presented cue and targets, or a result of slightly changing accommodation between the focus points of the cue and the target.

The inter-pupil distance variation was found to be significantly smaller (typically much less than $0.2\%$) than the recorded pupil dilations, and the ``baseline" experiment could not account for the recorded pupil dilations from the real experimental procedure either; it just showed the expected random variations.

The data processing has been done with iPython~\cite{perez2007ipython} using the numpy~\cite{vanderwalt2011numpy}, matplotlib~\cite{hunter2007matplotlib}, pandas~\cite{mckinney2010data}, scipy~\cite{oliphant2007scipy} and scikit-learn~\cite{pedregosa2012scikitlearn} toolboxes.

\section{Results}

\subsection{Attention Network Test timings}

Table \ref{tab:rtant} shows the aggregate Overall Mean Reaction- and Attention Network timings for each subject A and B, with estimates of the variation over the week. The figures are not significantly different from what is found in~\cite{fan2002testing}; the Mean\textsc{rt} reported here is slightly higher than an estimated $512$ms in the reference, whereas the alertness, orientation and conflict resolution are slightly lower or similar to the $47$ms, $51$ms and $84$ms reported.
\begin{table}[]
\centering
\caption{Average Reaction- and Attention Network-Times over all correctly replied experiments for the two week period for either user (the variation over the period is given as estimated $\pm$~Sample Standard Deviation of the aggregate values), in milliseconds.}
\label{tab:rtant}
\begin{tabularx}{0.75\textwidth}{@{}l*{4}{C}@{}}
\toprule
Subject & Mean\textsc{rt} & Alert & Orient & Conflict \\ \midrule
A & $577$ ($\pm54$) & $27$ ($\pm21$) & $22$ ($\pm18$) & $85$ ($\pm16$) \\
B & $559$ ($\pm55$) & $35$ ($\pm17$) & $49$ ($\pm15$) & $81$ ($\pm17$) \\ \bottomrule
\end{tabularx}
\end{table}


%
%

\begin{figure}[h!]
\includegraphics[width=0.5\textwidth, trim=12 0 36 24]{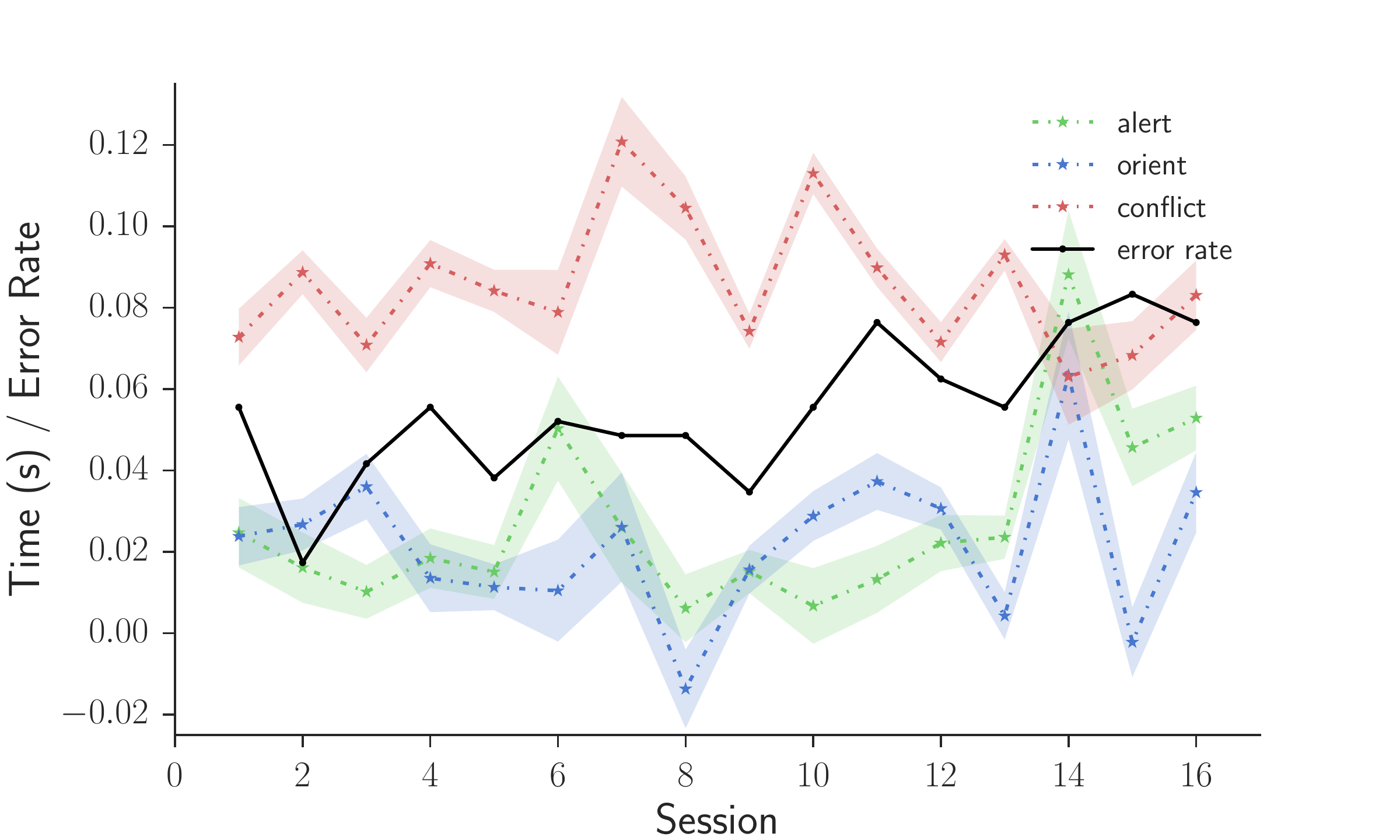}
\includegraphics[width=0.5\textwidth, trim=12 0 36 24]{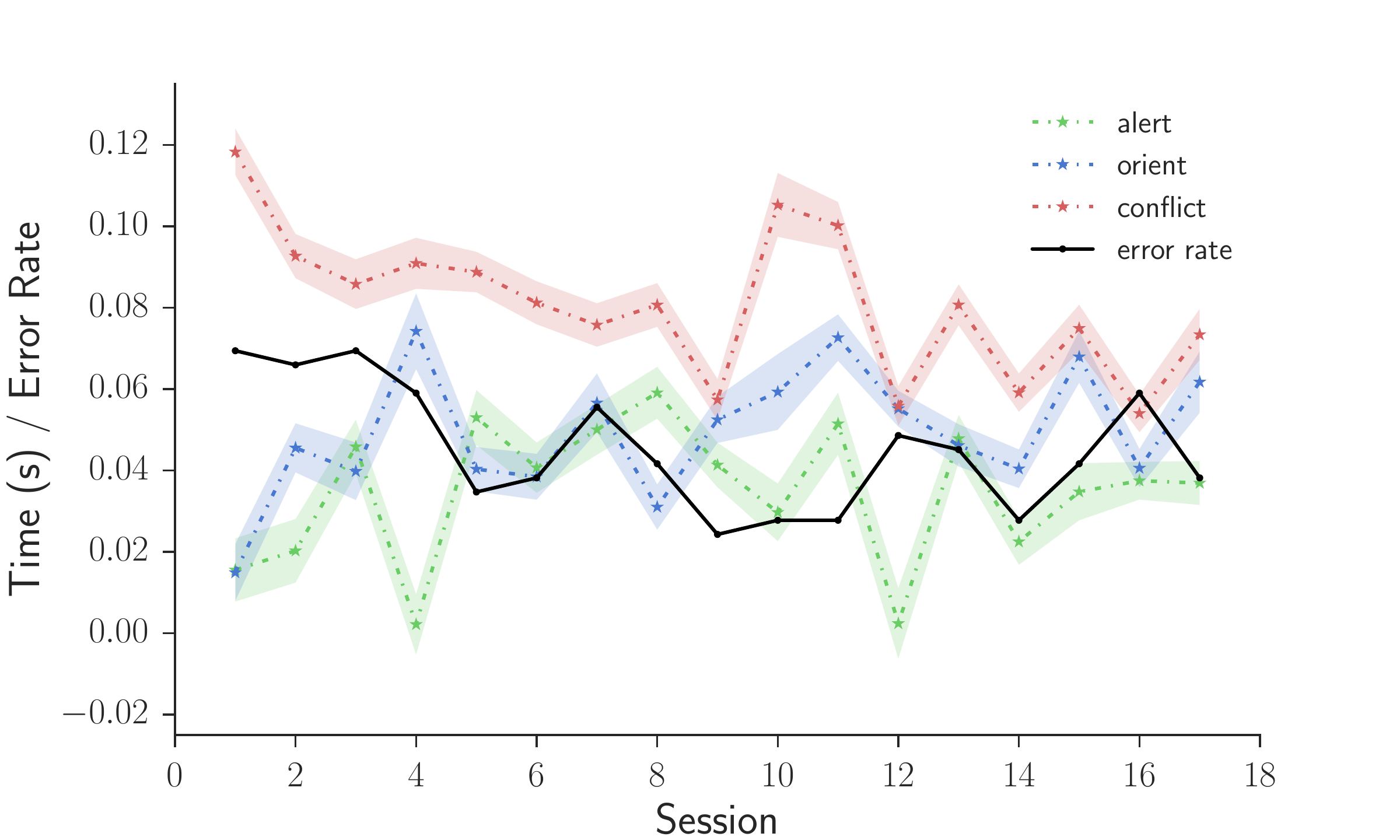}
\caption[LoF Entry]{
Attention Network Timing over all sessions in the two week period. 
Conflict Resolution (\textsc{Red}) is slower than Alertness (\textsc{Green}) and Orientation (\textsc{Blue}). 
A (\textsc{Left}) shows an increasing error rate trend (\textsc{Solid}); Conflict Resolution for B gradually approaches the other latencies.
Both A and B have large variations over time, pointing to varying levels of attention, fatigue and motivation.
}
\label{fig:ant}
\end{figure}

There are, however, behavioural variations in reaction time throughout the weeks. Fig.~\ref{fig:ant} shows the variation of the derived \textsc{ant} timings throughout the experimental period, and the relative error rate for each experiment. The variation appear to be statistically significant, as can be estimated from the standard error of the mean (the shaded area), and may reflect underlying states of varying levels of attention, fatigue and motivation.

To sum up the behavioral results, A shows a somewhat increasing trend in error rate related to the objective task performance, whereas B shows a diminishing difference between the three estimated measures of conflict resolution, spatial orientation and alertness reaction time.

\subsection{Pupil Dilations}

\begin{figure}[h!]
\includegraphics[width=0.5\textwidth, trim=24 0 36 24]{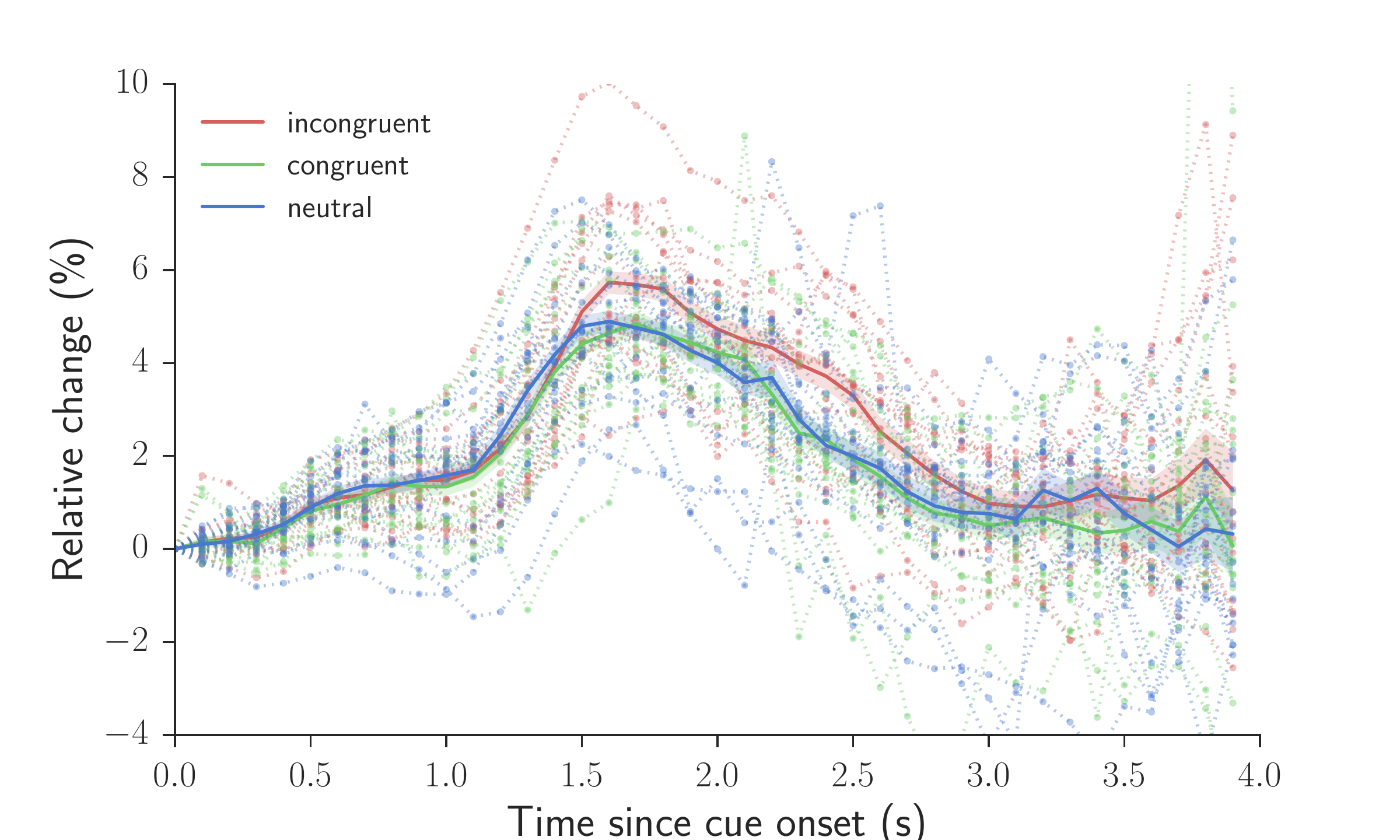}
\includegraphics[width=0.5\textwidth, trim=24 0 36 24]{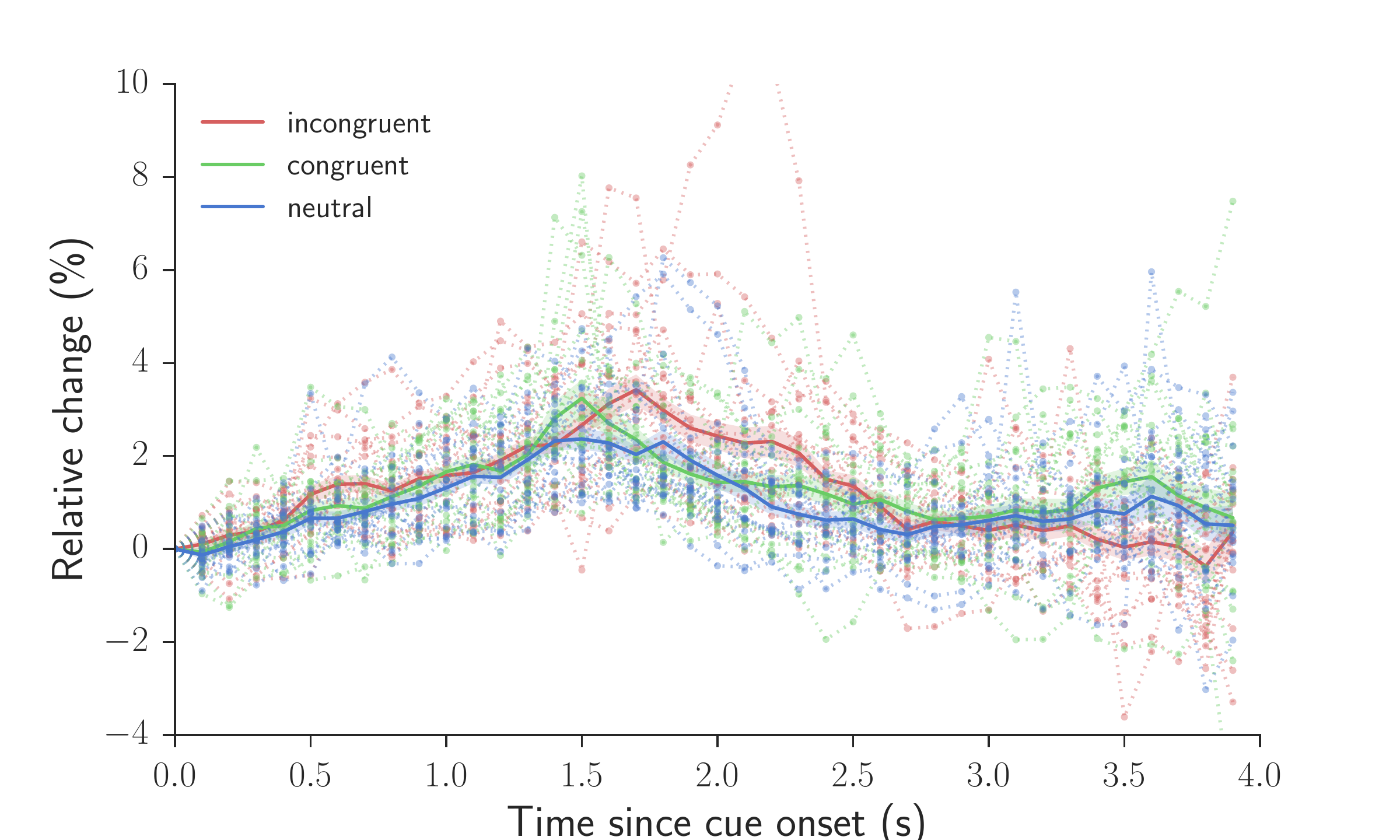}
\caption[LoF Entry]{
Averaged left-eye pupil dilations for each session, coloured according to congruency (A (\textsc{Left}) and B). All-session average shown in bold, with the shaded area representing the standard error of the mean.
The average incongruent (\textsc{Red}) pupil dilation is stronger than the others, indicating a higher cognitive load.
}
\label{fig:pupil-all}
\end{figure}

The group-mean relative linear pupil dilations for each of the 3 congruency conditions are illustrated in Fig.~\ref{fig:pupil-all}.

Pupil dilation responses are all epoch'ed to the cue (at time $0$ms) and target presentation (time $500$ms). A small and slow pupil dilation onset is seen $<300$ms after cue presentation, followed by a larger response likely triggered by the target presentation, with an onset of approximately $700$ms and a peak approximately $1300$ms after target, with some variation between conditions, subject and eye.


Even though the experimental conditions are not directly comparable, \cite{laeng2011pupillary} reported comparable peak latencies at $1400$ms after stimulus for a Stroop effect experiment. Our results are thus in line with these previous findings of pupil dilations, as well as with those reported in earlier processing load experiments \cite{hyona1995pupil} at approximately $900-1200$ms. The initial onset of the pupil dilation can occur even faster in some conditions \cite{beatty1982task} \cite{holmqvist2011eye} although generally onset and peak latencies appear to be within the $150-1400$ms.

The incongruent pupil dilation is larger than the more similar neutral and congruent dilations; there is however no such difference when comparing the 4 cue condition (not shown). The incongruent pupil dilation also has a tendency to appear slightly later (most easily visible for A), consistent with the longer reaction times for the inconsistent condition.

\begin{figure}[h!]
\includegraphics[width=0.5\textwidth, trim=24 0 36 24]{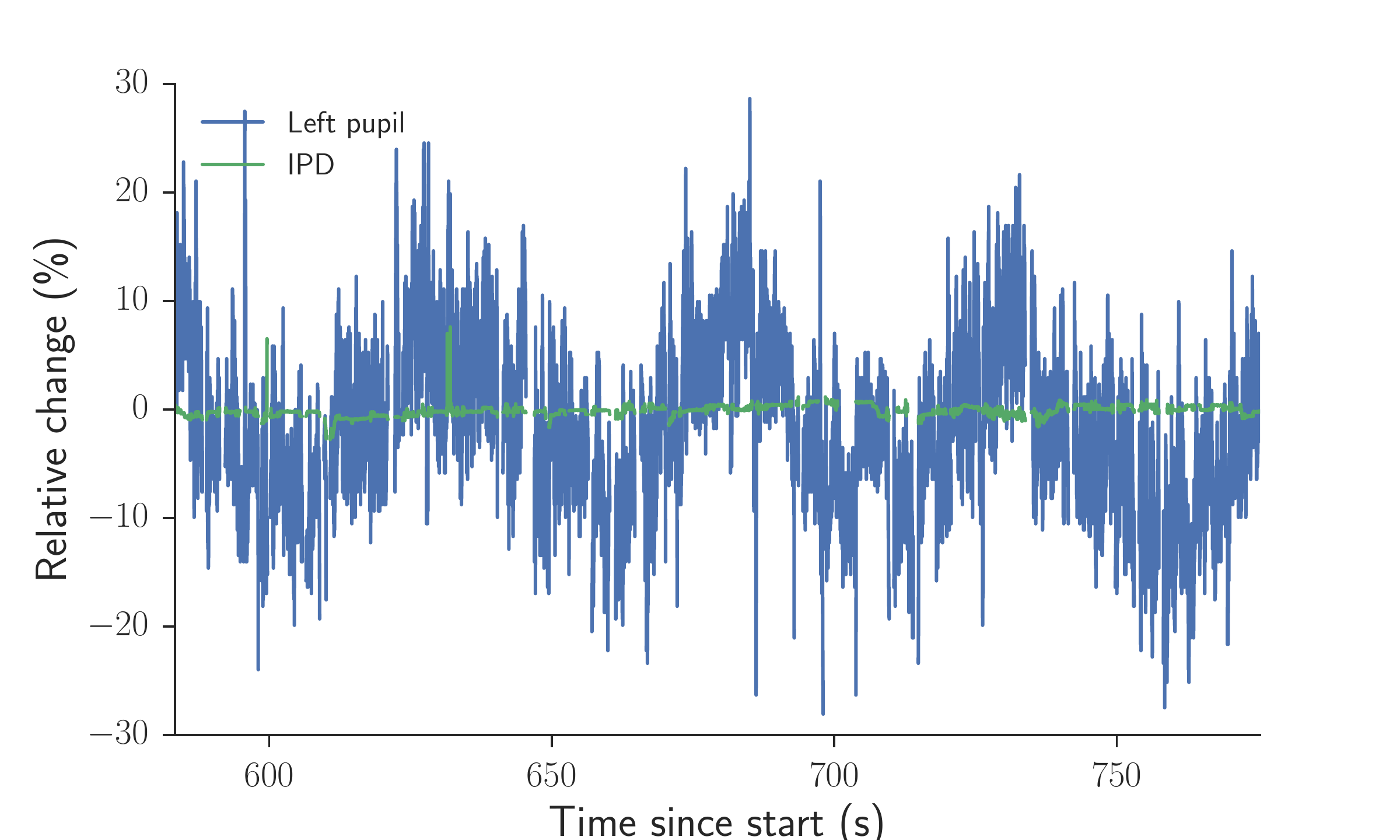}
\includegraphics[width=0.5\textwidth, trim=24 0 36 24]{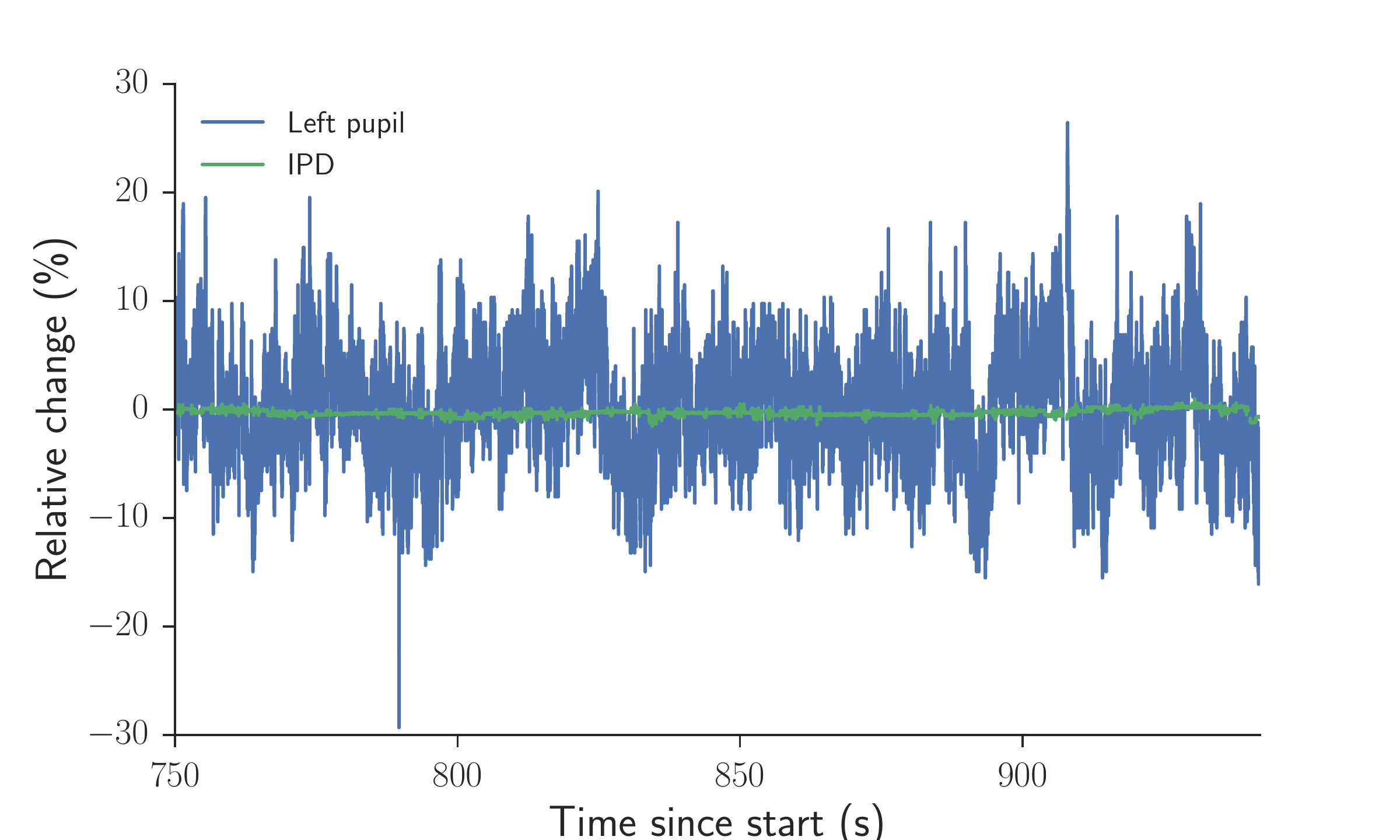}
\caption[LoF Entry]{
Filtered pupil size plots; 48-test long sections of two experiments (B, left-eye).  
Relative inter-pupil distance (\textsc{Green}) indicates stable eye-to-screen distances.
}
\label{fig:pupil-long}
\end{figure}

Fig~\ref{fig:pupil-long} shows the (relative) pupil size \textsc{Blue} vs the median value over a selected period that covers 48 reaction time tests, in this case for B, for two different experiments. Test-related pupil dilation responses, that occur every 4 seconds, are not immediately visible in this graph due to random noise and a relatively strong longer-periodic variation over 20-60 seconds\footnote{A frequency domain analysis of the signal shows, however, a distinct peak at $0.25$ Hz, as expected}. The \textsc{Green} curve shows the relative variation of the inter-pupil distance, with variations an order of magnitude smaller than the pupil size changes.

\begin{figure}[h!]
\includegraphics[width=0.5\textwidth, trim=24 0 36 24]{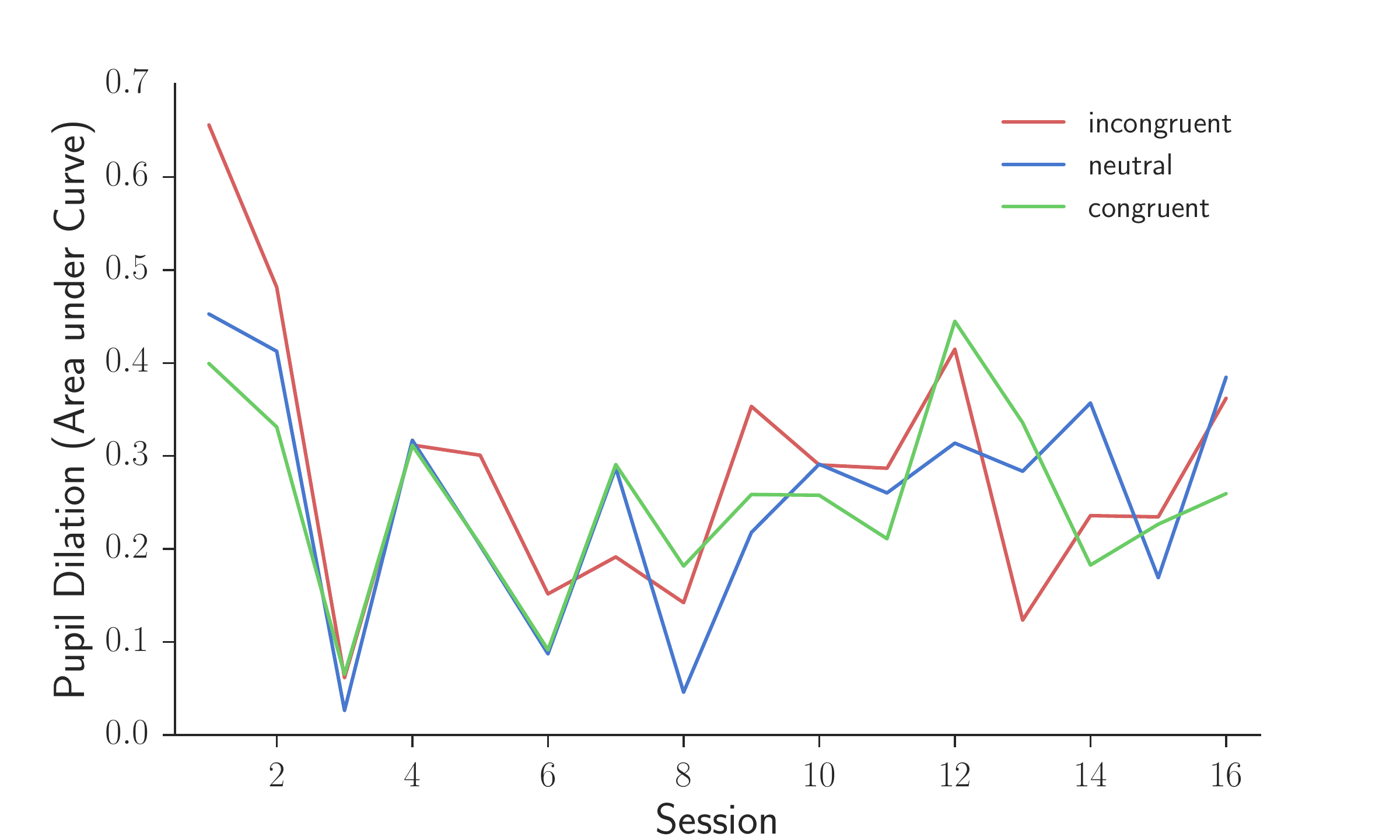}
\includegraphics[width=0.5\textwidth, trim=24 0 36 24]{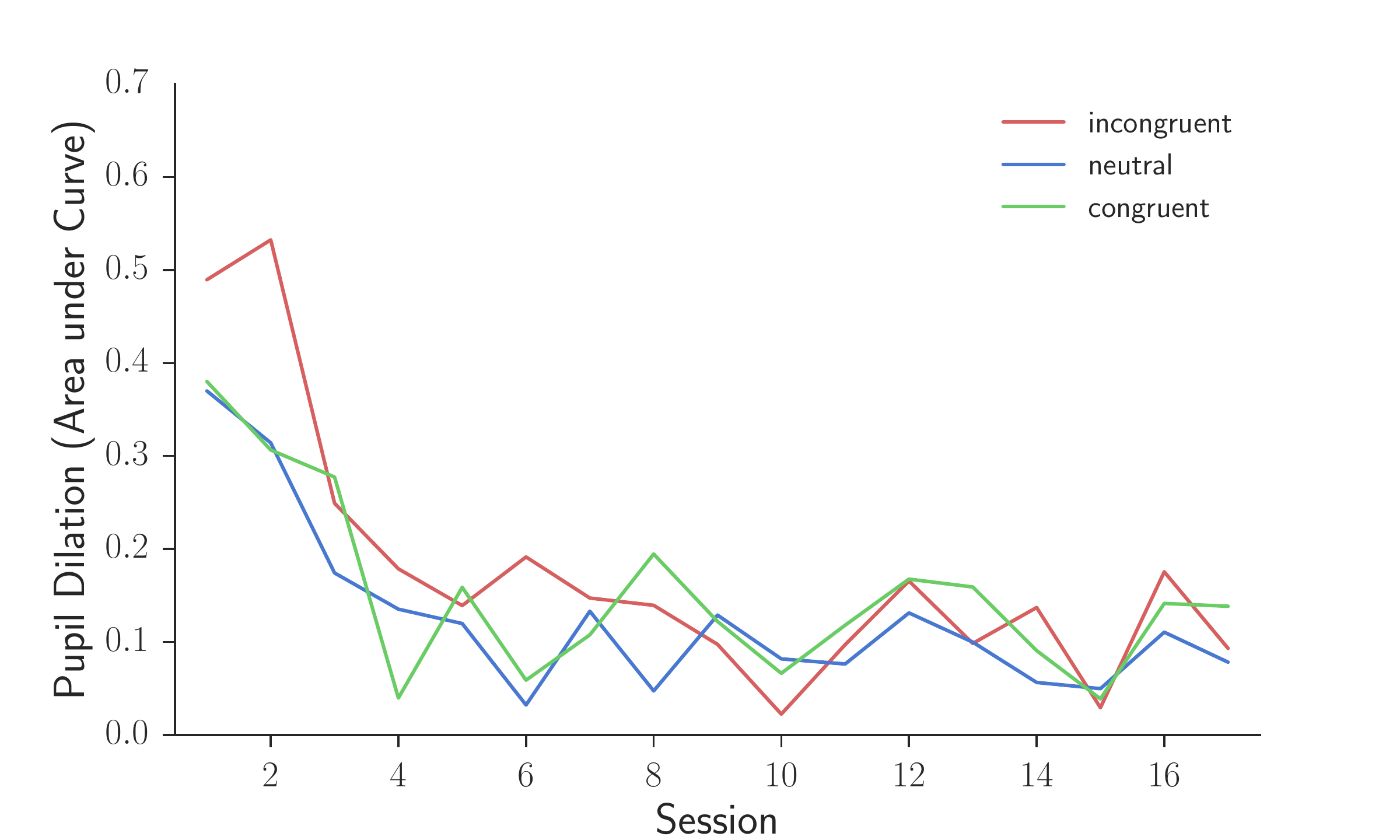}
\caption[LoF Entry]{
Area under left-eye pupil dilation curves $[1.5,2.5]$s for each session, indicative of cognitive load, grouped after congruency.
Both A (\textsc{Left}) and B show initial training effects; only A however shows an increasing trend in cognitive load for the remaining sessions.
}
\label{fig:area-vs-run}
\end{figure}

Fig~\ref{fig:area-vs-run} shows the area under the pupil dilation curve between $1.5-2.5$s after cue ($1.0-2.0$s after target) for each experiment, serving as a very rough indicator of the relative cognitive load caused by the tests. From these, also a $\delta$(incon) can be calculated by subtracting the congruent value from the incongruent.

It is seen that both A and B have larger pupil dilation responses for the initial two experiments, after which the level is lower. For B it remains at lower levels, indicating a training effect. For A, the pattern is less clear, with possibly an increased load towards the end of the two week period.

%
%

\subsection{Predicting congruency condition from pupil dilations}

In order to verify how well previous pupil dilations allow predicting the class of congruency condition, a subset of the $3$ within-experiment $96-$average pupil dilation responses from each subject were ordered in each of the 6 possible permutations of the 3 congruency conditions. A neural-network type classifier was then trained to identify which of the 3 averaged pupil dilations were the incongruent.

\begin{figure}[h!]
\centering{
\includegraphics[width=0.5\textwidth, trim=36 12 36 24]{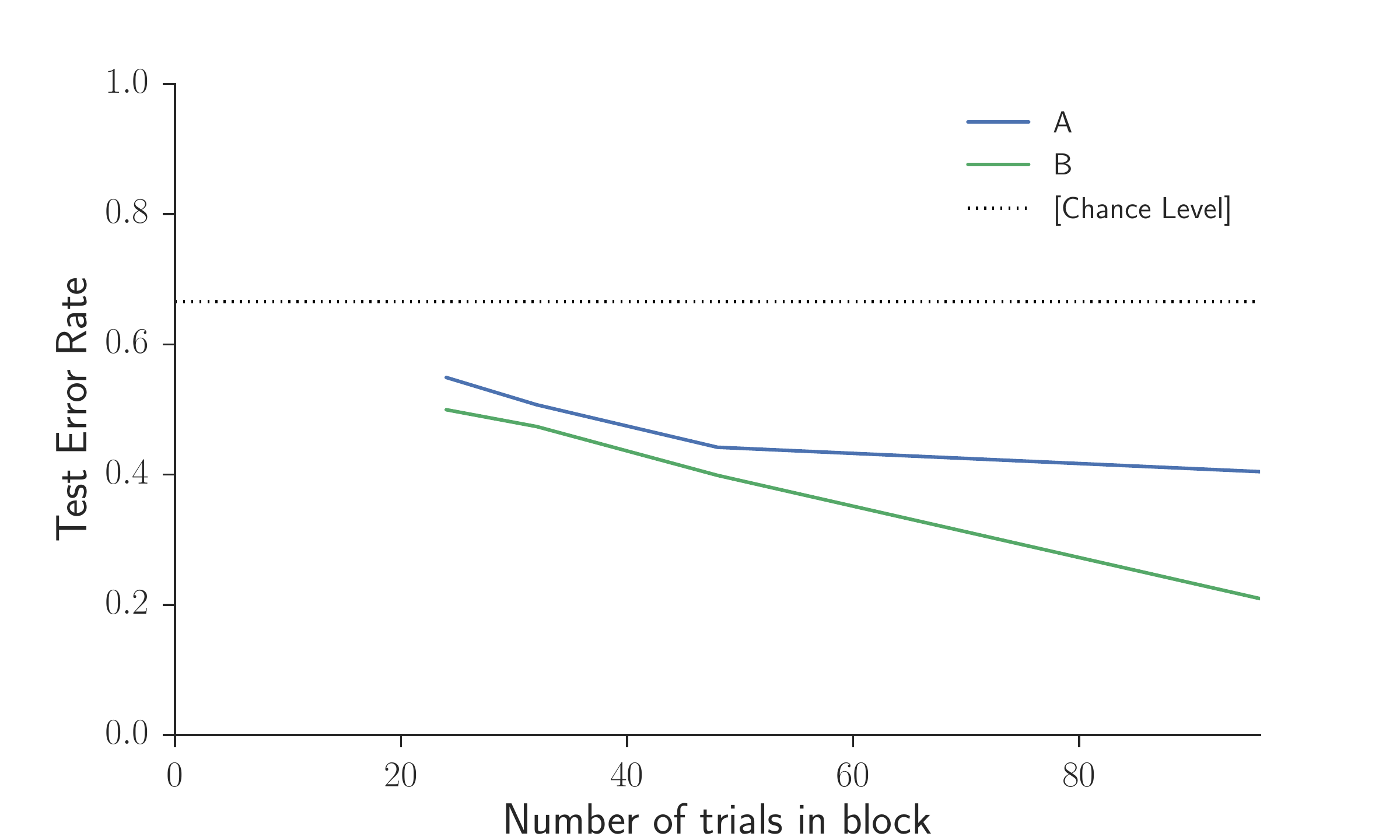}
}
\caption[LoF Entry]{
Test error rates ($0.9/0.1$ train/test split) predicting averaged $3$s incongruent pupil dilations after cue vs number of averaged experimental tests. 
At $48$ averaged experimental tests, the test error rate at $50\%$ is clearly below chance ($66.6\%$, dotted).
}
\label{fig:ml-predictability}
\end{figure}

Fig~\ref{fig:ml-predictability} shows the resulting test error rate vs. the number of averaged experimental tests, dividing the $96$ equal-condition responses of each experiment into groups of $96$, $48$, $32$ or $24$ tests, and using a test/train split of $0.9/0.1$. The performance is clearly above chance level ($66.6$\%), and approaches $80$\% accuracy for B vs $60$\% for A. Even at groups of $24$ averaged experimental tests, the classifier operate above chance level, with continuing improved performance for larger groups for B, however only marginally improving performance for A.

\subsection{Correlating response times and pupil reactions}

Table~\ref{tab:pearson-A} show the Pearson Correlation Coefficients for all combinations of Attention Network- and Reaction-Times, Pupil Dilation metrics and Time-of-Day for each subject, as it varies over the two week period. As the data sets are small (16 and 17 sets), caution is needed when judging the significance levels (p-values).

\begin{table}[]
\centering
\caption{
Pearsons correlation coefficients between key metrics for A (\textsc{Top}) and B.
A shows negative correlation between mean reaction time and error rate ("speed-accuracy tradeoff"). B (opposed to A) shows correlation between pupil dilations and error rate, possibly indicating a different response to varying levels of fatigue or motivation; additionally alertness (and partly orientation) may inversely correlate to pupil dilations.
Both show expected correlations between pupil dilation metrics.
}
\label{tab:pearson-A}
\scriptsize{

\begin{tabularx}{\textwidth}{l*{9}{R@{}l}}
\toprule
& \multicolumn{5}{c}{\emph{Att.-Net/Reaction Time}} & \phantom{i} & \multicolumn{7}{c}{\emph{Pupil Dilation}} & \phantom{i} & \\
\cmidrule{2-6} \cmidrule{8-14} \cmidrule{16-19}
& \multicolumn{1}{c}{Orient}&& \multicolumn{1}{c}{Conflict}&& \multicolumn{1}{c}{$\mu$(\textsc{RT})}&& \multicolumn{1}{c}{Incon}&& \multicolumn{1}{c}{Neutral}&& \multicolumn{1}{c}{Con}&& \multicolumn{1}{c}{$\delta$(Incon)}&& \multicolumn{1}{c}{\textsc{ToD}}&& \multicolumn{1}{c}{Errors} \\ \midrule
\multicolumn{8}{l}{\emph{Att.-Net/Reaction Time}}\\
Alert &$0.112$& &$-0.047$& &$-0.189$& &$-0.013$& &$-0.131$& &$-0.011$& &$-0.008$& &$0.061$& &$-0.051$&\\
Orient&&  &$\mathbf{-0.548}$&$^\dagger$ &$-0.468$&$^*$ &$0.274$& &$0.269$& &$-0.020$& &$0.402$& &$0.132$& &$0.270$&\\
Conflict&& &&  &$0.474$&$^*$ &$-0.081$& &$-0.149$& &$0.035$& &$-0.147$& &$0.330$& &$-0.416$&\\
$\mu$(\textsc{RT})&& && &&  &$0.002$& &$0.049$& &$-0.069$& &$0.068$& &$0.237$& &$\mathbf{-0.635}$&$^\dagger$\\
\multicolumn{8}{l}{\emph{Pupil Dilation}}\\
Incon&& && && &&  &$\mathbf{0.767}$&$^\ddagger$ &$\mathbf{0.701}$&$^\ddagger$ &$\mathbf{0.737}$&$^\ddagger$ &$0.062$& &$-0.098$&\\
Neutral&& && && && &&  &$\mathbf{0.752}$&$^\ddagger$ &$0.362$& &$0.222$& &$0.109$&\\
Con&& && && && && &&  &$0.034$& &$0.000$& &$-0.018$&\\
$\delta$(Incon)&& && && && && && &&  &$0.087$& &$-0.121$&\\
\textsc{ToD}&& && && && && && && &&  &$0.066$&\\
\bottomrule
\multicolumn{19}{c}{Two-tailed significance less than $^*7.5\%$, $\mathbf{^\dagger5\%}$ and $\mathbf{^\ddagger0.25\%}$ marked.}
\end{tabularx}

}
\vspace{0.5pc}
\scriptsize{

\begin{tabularx}{\textwidth}{l*{9}{R@{}l}}
\toprule
& \multicolumn{5}{c}{\emph{Att.-Net/Reaction Time}} & \phantom{i} & \multicolumn{7}{c}{\emph{Pupil Dilation}} & \phantom{i} & \\
\cmidrule{2-6} \cmidrule{8-14} \cmidrule{16-19}
& \multicolumn{1}{c}{Orient}&& \multicolumn{1}{c}{Conflict}&& \multicolumn{1}{c}{$\mu$(\textsc{RT})}&& \multicolumn{1}{c}{Incon}&& \multicolumn{1}{c}{Neutral}&& \multicolumn{1}{c}{Con}&& \multicolumn{1}{c}{$\delta$(Incon)}&& \multicolumn{1}{c}{\textsc{ToD}}&& \multicolumn{1}{c}{Errors} \\ \midrule
\multicolumn{8}{l}{\emph{Att.-Net/Reaction Time}}\\
Alert &$0.015$& &$-0.107$& &$0.438$& &$\mathbf{-0.499}$&$^\dagger$ &$\mathbf{-0.534}$&$^\dagger$ &$-0.231$& &$\mathbf{-0.576}$&$^\dagger$ &$0.062$& &$-0.358$&\\
Orient&&  &$-0.094$& &$0.352$& &$-0.474$&$^*$ &$-0.407$& &$\mathbf{-0.559}$&$^\dagger$ &$-0.155$& &$0.056$& &$-0.386$&\\
Conflict&& &&  &$0.289$& &$0.431$& &$0.439$& &$0.362$& &$0.309$& &$0.411$& &$0.301$&\\
$\mu$(\textsc{RT})&& && &&  &$-0.220$& &$-0.286$& &$-0.173$& &$-0.173$& &$0.481$&$^*$ &$-0.400$&\\
\multicolumn{8}{l}{\emph{Pupil Dilation}}\\
Incon&& && && &&  &$\mathbf{0.894}$&$^\ddagger$ &$\mathbf{0.817}$&$^\ddagger$ &$\mathbf{0.746}$&$^\ddagger$ &$-0.026$& &$\mathbf{0.725}$&$^\ddagger$\\
Neutral&& && && && &&  &$\mathbf{0.831}$&$^\ddagger$ &$\mathbf{0.549}$&$^\dagger$ &$-0.184$& &$\mathbf{0.701}$&$^\ddagger$\\
Con&& && && && && &&  &$0.224$& &$-0.020$& &$\mathbf{0.626}$&$^\dagger$\\
$\delta$(Incon)&& && && && && && &&  &$-0.021$& &$\mathbf{0.501}$&$^\dagger$\\
\textsc{ToD}&& && && && && && && &&  &$-0.215$&\\
\bottomrule
\multicolumn{19}{c}{Two-tailed significance less than $^*7.5\%$, $\mathbf{^\dagger5\%}$ and $\mathbf{^\ddagger0.25\%}$ marked.}
\end{tabularx}
 
}
\label{tab:pearson-B}
\end{table}

With some variation between subjects, pupil dilation responses appear correlated.


Subject A shows correlation between orientation and conflict resolution timings, which is however not seen at all for B. A also may have some correlation between mean reaction time and orientation resp conflict resolution timings, which are however again not quite as present with B.

Subject B shows correlation between alertness timing and both incongruent, neutral and $\delta$(incon) pupil dilations, as well as correlation between orientation timing and congruent pupil dilations. These are not present for A, however. Also, there are indications of a correlation between the time of day and the mean reaction time; the experiments done on B were spread out over larger sections of the day than for A, which might explain why this is not seen for A.

\cite{fan2002testing} reported correlations between the conflict resolution timing and the mean reaction time over a large group of people. As such, the conditions are not similar to the within-person variation, but it might be worth pointing out that a similar correlation is partly present for A and cannot be ruled out for B.

\section{Discussion}

Using low cost portable eye tracking to measure the variations in pupil size, we were able to differentiate and predict whether users were engaged in more complex decision making or merely maintaining a general alertness when interacting with a laptop, over nearly $10.000$ tests. 
A parallel single-experiment study \cite{baekgaard2016notitle} repeating the experimental setup with nearly $10.000$ additional tests over 18 more subjects, have confirmed that similar significant pupil response differences characterize the contrasts between incongruent versus neutral or congruent task conditions.

In the present study, we found a significant difference based on the left eye pupil size for the conflict resolution task in contrast to the attentional network components of alertness and re-orientation, but not between these two latter tasks. These results may reflect findings in other studies indicating that the phasic component in attention is predominantly triggered by tasks requiring a decision, whereas the tonic alertness may suffice for solving less demanding tasks like responding to visual cues or re-orienting attention to an unexpected part of the screen~\cite{aston-jones2005integrative} as seen in the ``baseline" experiment, where no decision needs to be made and no motor cortex activation takes place.


From a quantified self perspective of individual behaviour, using mobile eye tracking to assess levels of engagement, the relations between pupil size (a possible quantification of the cognitive load), and error rate/reaction time (a quantification of the objective task performance), indicate individual differences among the subjects' behavioural adaptation to the attentional tasks. A is apparently coping with the cognitive load by trading off speed and accuracy to optimize performance, as indicated by the lack of correlation between pupil size and either of the performance related measures. However, for B the correlation between pupil size and accuracy may suggest a behavior characterized by applying more effort to the task if the number of errors increase.

As we have in this study only used the pupil size as a measure of attention, even without considering the spatial density of fixations or the speed of saccadic eye movements that could entail further information, we suggest that mobile eye tracking may not only enable us to assess the effort required when undertaking a variety of tasks in an everyday context, but could also longer term provide a foundation for continuously adapting the content and interaction with smartphones and laptops based on our perceived level of attention.

\bibliography{baekgaard2016assessing}
%
\end{document}